\documentclass[reprint,twocolumn,prb,floatfix,superscriptaddress,showpacs,citeautoscript]{revtex4}
\usepackage{graphicx}
\usepackage{amssymb}
\usepackage{pifont}
\usepackage{dcolumn}

\begin{document}

\title{Detection and quantification of inverse spin Hall effect from spin pumping in permalloy/normal metal bilayers}

\author{O.~Mosendz}
\altaffiliation{New address: San Jose Research Center, Hitachi Global Storage Technologies, San Jose, California 95135, USA}
\affiliation{Materials Science Division,
Argonne National Laboratory, Argonne, IL 60439, USA}

\author{V.~Vlaminck}
\affiliation{Materials Science Division, Argonne National
Laboratory, Argonne, IL 60439, USA}

\author{J.~E.~Pearson}
\affiliation{Materials Science Division, Argonne National
Laboratory, Argonne, IL 60439, USA}

\author{F.~Y.~Fradin}
\affiliation{Materials Science Division, Argonne National
Laboratory, Argonne, IL 60439, USA}

\author{G.~E.~W.~Bauer}
\affiliation{Kavli Institute of NanoScience, Delft University of
Technology, 2628 CJ Delft, The Netherlands}

\author{S.~D.~Bader}
\affiliation{Materials Science Division, Argonne National
Laboratory, Argonne, IL 60439, USA} \affiliation{Center for
Nanoscale Materials, Argonne National Laboratory, Argonne, IL 60439,
USA}

\author{A.~Hoffmann}
\email{hoffmann@anl.gov}\affiliation{Materials Science Division, Argonne National
Laboratory, Argonne, IL 60439, USA}



\date{\today}

\begin{abstract}
Spin pumping is a mechanism that generates spin currents from ferromagnetic resonance (FMR) over macroscopic interfacial areas, thereby enabling sensitive detection of the inverse spin Hall effect that transforms spin into charge currents in non-magnetic conductors.  Here we study the spin-pumping-induced voltages due to the inverse spin Hall effect in permalloy/normal metal bilayers integrated into coplanar waveguides for different normal metals and as a function of angle of the applied magnetic field direction, as well as microwave frequency and power.  We find good agreement between experimental data and a theoretical model that includes contributions from anisotropic magnetoresistance (AMR) and inverse spin Hall effect (ISHE).  The analysis provides consistent results over a wide range of experimental conditions as long as the precise magnetization trajectory is taken into account.  The spin Hall angles for Pt, Pd, Au and Mo were determined with high precision to be $0.013\pm0.002$, $0.0064\pm0.001$, $0.0035\pm0.0003$ and $-0.0005\pm0.0001$, respectively.
\end{abstract}



\maketitle


\section{Introduction}
Information in semiconductor electronic devices and data storage technologies is mainly
transported and manipulated by charge currents. With advancing miniaturization, heat dissipation and power consumption become significant obstacles to further technological advances. Alternative technologies that solve or at least circumvent these problems are needed.  One promising candidate to replace existing charge-based technologies is based on using spin currents; an effort referred to as spintronics\cite{Bader-ARCMP2010}. Magnetoelectronic devices employing spin-polarized charge currents are already actively in use in hard drive read-heads and non-volatile magnetic random access memories (MRAMs).

{\em Pure} spin currents that are not accompanied by a net charge current, may offer additional advantages in applications,\cite{Chappert-NP2008,Hoffmann-PSSC2007} such as reduced power dissipation, absence of stray Oersted fields, and decoupling of spin and charge noise.  Furthermore, undisturbed by charge transport, pure spin currents can provide more direct insights into the basic physics of spin-dependent effects.  Pure spin currents can be created by, for instance: (i) Non-local electrical injection from ferromagnetic contacts in multi-terminal structures; (ii) optical injection using circularly polarized light; (iii) spin pumping from a precessing ferromagnet; and (iv) spin Hall effect.  The last possibility is particularly interesting, since ferromagnets are not 
involved\cite{Dyakonov-JETF,Hirsch,Zhang-PRL2000}.  The spin Hall effect is caused by the spin-orbit interactions of defect scattering potentials or the host electronic structure.  The efficiency of this spin-charge conversion can be quantified by a single material-specific parameter, {\em viz.}\ the spin Hall angle $\gamma$, which is defined as the ratio of the spin Hall and charge 
conductivities \cite{Dyakonov-BookChapter2008} and can be measured by magnetotransport measurements 
\cite{Fert-JMMM1981,Valenzuela-Nature2006,Kimura-PRL2007,Seki-NM2008,Morota-JAP09}.
Previous experimental studies report quite different $\gamma$ values for nominally identical materials.  For example, for Au a giant $\gamma = 0.113$ was reported \cite{Seki-NM2008}, while subsequent experiments found values that are one or even two orders of magnitude smaller\cite{Mihajlovic-PRL2009,Mosendz-PRL2010}.  Similarly, for Pt different experiments \cite{Kimura-PRL2007,Mosendz-PRL2010,Ando-PRL2008} resulted in $\gamma$ values that vary between 0.0037 and 0.08.

Recently we demonstrated a robust technique \cite{Mosendz-PRL2010} to measure spin Hall angles with high accuracy in arbitrary conductors.  Our approach is based on the combination of spin pumping, which generates pure spin currents, and measurements of electric voltages due to the inverse spin Hall effect (ISHE)\cite{Saitoh-APL06}.  Here we present a detailed discussion of the measurements in Ref.~\onlinecite{Mosendz-PRL2010} and examine the validity of the theoretical model used to describe the voltages induced in the Ni$_{80}$Fe$_{20}$(Py)/normal metal (N) bilayers.  In particular we measure the ISHE voltage as a function of angle of the applied magnetic field, and microwave frequency and power. We find excellent agreement between model calculations and experimental results.  Accounting for the proper magnetization trajectory is important for a quantitative interpretation of the results.  Good agreement between the theoretical model and experiments for a wide range of controlled experimental parameters implies that our approach is robust and can be used to determine the magnitude and sign of spin Hall effects in more conductors than included in the present study.

\section{Coupling between spin and charge currents}
\subsection{Spin pumping in Py/N bilayers}
\begin{figure}%
    \includegraphics[width=8.6cm]{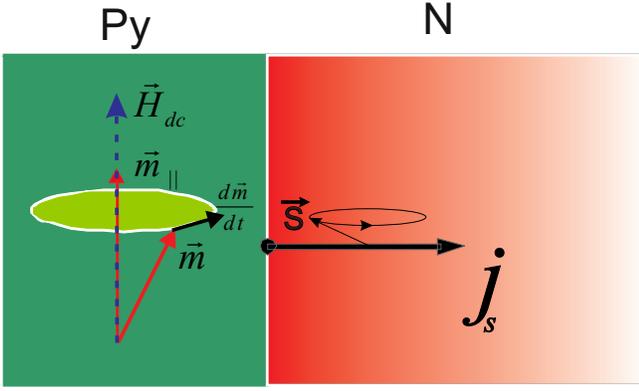}
    \caption{(Color online) Schematic model of spin pumping in Py/N bilayer. $\vec{s}$ shows that the polarization of the spin current is oscillating in time with the frequency of the magnetization precession. The polarization of the spin current is perpendicular to the instantaneous magnetization direction $\vec{m}$, and the rate of magnetization change 
 $\partial \vec{m} / \partial t$. 
    } %
    \label{spinpump}%
\end{figure}

Spin pumping generates pure spin currents in normal metals (N), when they are in contact with a ferromagnet with time-dependent magnetization induced, {\em e.g.}, by ferromagnetic resonance (FMR)\cite{Heinrich-PRL2003,Woltersdorf-PRL2007,Mosendz-PRB09,Kardasz-JAP2008,Kardasz-PRB2010}.  The instantaneous spin-pumping current $j_s^0$ at the Py/N interface is given 
by:\cite{Brataas-PRL2002,Yaroslav-review}
\begin{equation}\label{spincurrent}
j_s^0\vec{s}=\frac{\hbar}{8\pi}Re
(2g^{\uparrow\downarrow})\left[\vec{m}\times \frac{\partial\vec{m}}{\partial t}\right],
\end{equation}
where $\vec{m}$ is the unit vector of the magnetization, $\vec{s}$ is the unit vector of the spin current polarization in N, and $Re(g^{\uparrow\downarrow})$ is the real part of the spin-mixing
conductance. The spin current generated by spin pumping is polarized perpendicular to the instantaneous magnetization direction $\vec{m}$ and its time derivative 
$\partial\vec{m}/\partial t$ (see Fig.~\ref{spinpump}). Note that this spin current always has a polarization component along $\vec{ H}_{dc}$ and propagates into N normal to the interface.

The spin current generated at the Py/N interface accumulates a spin density $\vec{\mu}_N$ inside the N layer.  In the ballistic limit ({\em i.e.}, no spin relaxation in N) the spin current reaching the N/vacuum interface is fully reflected and reabsorbed upon returning to the Py/N interface, without influencing the magnetization dynamics of the bilayer system.  

In real systems, pure spin currents are not conserved, since spins relax over length scales given by the spin diffusion length $\lambda_{sd}$ in N, and the accumulated spin density moves across the N layer via spin diffusion limited by momentum scattering (leading to electrical resistance) and spin-flip scattering (leading to loss of spin angular momentum) by spin orbit coupling or magnetic impurities.  The spin diffusion equation describes the dissipative propagation of the spin accumulation (difference in local electrochemical potentials of up and down spins) 
$\vec{\mu}_N$ in the N layer:
\begin{equation}\label{diffusion}
   i\omega\vec{\mu}_N = D\frac{\partial^{2}\vec{\mu}_N}{\partial z^{2}}-\frac{1}
   {\tau_{sf}}\vec{\mu}_N,
\end{equation}
where $\omega$ is the angular frequency, $\tau_{sf}$ is the spin-flip time, $z$ is the coordinate normal to the interface, and $D = v_{F}^{2}\tau_{el}/3$ is the electron diffusion constant, with 
$\tau_{el}$ the electron momentum relaxation time\cite{Yaroslav-2002}. The solutions of 
Eq.~(\ref{diffusion}) depend on the boundary conditions.  For a single magnetic layer structure Py/N the boundary condition at the Py/N interface is given by~\cite{Yaroslav-review}
\begin{equation}\label{boundary1}
j_s^0\vec{s}(z=0) = -D\frac{\partial\vec{\mu}_N}{\partial z }\Bigm\vert_{z=0},
\end{equation}
while for the outer interface we use the free magnetic moment
condition (full spin current reflection)
\begin{equation}\label{boundary2}
\frac{\partial\vec{\mu}_N}{\partial z }\Bigm\vert_{z=L}=0.
\end{equation}
Equations~(\ref{diffusion})--(\ref{boundary2}) can be solved analytically to yield the decay of the spin accumulation as a function of the distance from the Py/N interface. This decay results in spin accumulation profile in the N layer, which decays as a function of the distance from the interface, thus driving a spin current with a {\em dc} contribution:
\begin{equation}\label{zdepspincurrent}
j_{s}(z) = j_{s}^{0}\frac{\sinh
[(z-t_{N})/\lambda_{sd}]}{\sinh (t_{N}/\lambda_{sd})},
\end{equation}
where $t_{N}$ is the thickness of the N layer.  The spin accumulation in N gives rise to spin backflow into the ferromagnet, which effectively reduces the spin pumping current, which can be accounted for by replacing $g^{\uparrow\downarrow}$ in Eq.~(\ref{spincurrent}) with an effective spin mixing conductance $g^{\uparrow\downarrow}_{eff}$.\cite{Yaroslav-2002}

In FMR experiments the absorption of the microwave field that excites the magnetization is monitored.  The magnetization dynamics in ferromagnetic films can be described by the Landau-Lifshitz-Gilbert (LLG) equation of motion:
\begin{equation}\label{Brataas1}
 \frac{1}{\gamma_{g}}\frac{\partial \vec{m}}{\partial t}=-\left[\vec{m}\times \vec{H}_{eff}\right]+
 \frac{\alpha_G}{\gamma_{g}} \left[\vec{m}\times \frac{\partial \vec{m}}{\partial t}\right] ,
\end{equation}
where $\gamma_{g}=g e /2mc$ is the absolute value of the gyromagnetic ratio, and $\alpha_G$ is the dimensionless Gilbert damping parameter. The first term on the right-hand side represents the precessional torque due to the effective internal field $\vec{H}_{eff}$, which for the case of permalloy with small anisotropy is approximately equal to the externally applied magnetic field 
$H_{dc}$.  The second term in Eq.~(\ref{Brataas1}) represents the Gilbert damping 
torque~\cite{Heinrich-AinP93,Heinrich-III}. The spin pumping can be accounted for in the LLG equation of motion by adding a spin pumping contribution $\alpha_{sp}$ to $\alpha_G$, {\em i.e.}, the effective damping becomes $\alpha_{eff}=\alpha_G+\alpha_{sp}$.  The damping can be quantified by measuring the FMR line width $\Delta H$, half width at half maximum (HWHM), of the imaginary part of the {\em rf} susceptibility $\chi^{''}$, which is commonly measured at a constant microwave frequency by sweeping the {\em dc} magnetic field $H_{dc}$.  In case of Gilbert damping, $\Delta H$ depends linearly on the microwave angular frequency $\omega_f$, {\em i.e.}, 
$\Delta H = \alpha_{eff}\omega_f/\gamma_g$. The difference in the damping parameter, determined by the FMR line width, for samples without capping layer and samples in which the capping N layer is sufficiently thick to fully dissipate the pumped magnetic moment, is attributed to the loss of spin momentum in Py due to relaxation of the spin accumulation in N. This permits the
determination of the additional interface damping due to spin pumping~\cite{Urban-PRL01}, which in turn fixes the interfacial spin-mixing conductance to:
\begin{equation}
  g^{\uparrow\downarrow}_{eff}=\frac{4\pi\gamma_{g}M_{s}t_{Py}}{g\mu_{B}\omega_f}(\Delta H_{Py\mid N}-\Delta H_{Py}) \label{spinmix} ,
\end{equation}
where $t_{Py}$ is the Py layer thickness, $M_s$ is the Py saturation magnetization, and $\mu_{B}$ is the Bohr magneton.  Note that Eq.~(\ref{spinmix}) is only applicable when the damping is governed by the Gilbert phenomenology or $\Delta H \propto \omega_f$, {\em i.e.}, when inhomogeneous linewidth broadening is negligible.  Otherwise $g^{\uparrow\downarrow}_{eff}$ can still be determined from the additional Gilbert-like damping contribution $\alpha_{sp} = \alpha_{Py\mid N}-\alpha_{Py}$, where the latter two are obtained from the linewidth difference that scales linear with frequency, i.e., $\Delta H = \Delta H_{ih}+\alpha_{eff}\omega_f/\gamma_g$, where $\Delta H_{ih}$ is the sample-dependent inhomogeneous linewidth, measured as the zero-frequency intercept.

The  {\em dc} component of the spin current pumped into N is polarized parallel to the equilibrium magnetization and has previously been detected via a {\em dc} voltage {\em normal} to the Py/N interface~\cite{Costache-PRL06}. Under a simple circular precession of the Py magnetization the time-averaging of the spin current from Eq.~(\ref{spincurrent}) for small precession cone angles $\theta$ reads:
\begin{equation}\label{dcspincurrent}
j_{s,dc}^{0,circ}=\frac{\hbar\omega_f}{4\pi}Re g^{\uparrow\downarrow}_{eff}\sin^{2}\theta.
\end{equation}
In thin magnetic films the trajectory of the magnetization precession is not circular but elliptic due to the strong demagnetizing fields, which force the magnetization into the film-plane. The time-dependent cone angle $\theta$ modifies the \textit{dc} component of the pumped spin current by an ellipticity correction factor $P$ as derived and measured by Ando {\em et al.}.\cite{Ando-APL09}  For an in-plane equilibrium magnetization $j_{s,dc}^{eff}=P\ast j_{s,dc}^{circ}$ with:
\begin{equation}\label{corrfacelll}
P=\frac{2\omega_f\left[\gamma_{g}4\pi M_{s}+\sqrt{(\gamma_{g}4\pi M_{s})^{2}+
(2\omega_f)^{2}}\right]}{(\gamma_{g}4\pi M_{s})^{2}+(2\omega_f)^{2}}.
\end{equation}
Equation~(\ref{corrfacelll}) is a non-monotonic function of $\omega_f$, and $P$ can become slightly larger than 1, but tends towards 1 for high frequencies, {\em i.e.}, large applied fields.
\subsection{Inverse spin Hall effect}
Spin-orbit coupling or magnetic impurities give rise to different scattering directions for electrons with opposite spin. In their presence, a spin current in N induces a transverse Hall voltage.  This ISHE transforms spin currents into electrical voltage differences over the sample edges.  Spin pumping generates {\em dc} and {\em ac} components to the spin current: $j_{s,dc}^{eff}$ and an {\em rf} component transverse to the equilibrium magnetization direction. In this paper we address only the ISHE effect generated by the {\em dc} component $j_{s,dc}^{eff}$.

The {\em dc} ISHE transverse charge current reads:
\begin{equation}
\vec{j}_{c}^{ISH}(z) = \gamma (2e/\hbar)j_{s,dc}^{eff}[\vec{n} \times \langle\vec{s}\rangle],
\end{equation}
where $\gamma$ is the spin Hall angle, $\vec{n}$ is the unit vector normal to the interface and 
$\langle\vec{s}\rangle$ is the polarization vector of the \textit{dc} spin current. For $j_{s,dc}^{eff}$ the spin-polarization $\langle\vec{s}\rangle$ is along the equilibrium magnetization direction in Py. The \textit{dc} electric field lies in N in the plane of the films and perpendicular to the equilibrium magnetization of Py\cite{Mosendz-PRL2010,Saitoh-APL06,Saitoh-PRB08}.

\section{Experimental results}
Here we elucidate our method to obtain voltage signals due to the ISHE in various Py/N combinations under FMR conditions, thereby determining the spin Hall angle $\gamma$ with high accuracy.  The measured voltage signals scale with the sample length and, therefore, can be increased readily by making the samples longer.  We identify two contributions to the {\em dc} voltage:  one stems from the anisotropic magnetoresistance (AMR) and the second from the ISHE, which can be distinguished by their symmetries with respect to the field-offset from the resonance field.  Furthermore, we present a theoretical model for the spin Hall angle contribution and test its functional dependence of several parameters that can be controlled experimentally.

\subsection{Experimental set-up}
\begin{figure}
  \includegraphics[width=8.6cm]{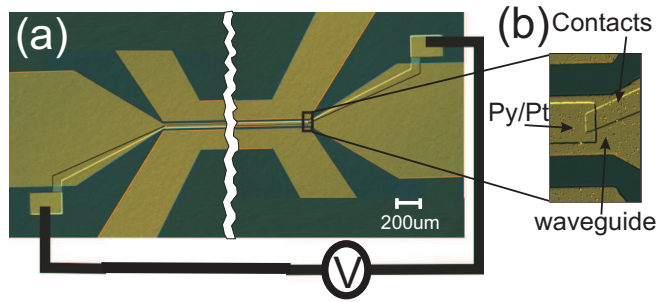}
  \caption{(Color online) Experimental setup:  (a) Optical image of the Py/Pt bilayer integrated into the coplanar waveguide. (b) Contacts are added at the end of the bilayer to measure the voltage along the waveguide direction.} %
  \label{setup1}%
\end{figure}
The Py/N bilayers were integrated into coplanar waveguides with additional leads in order to measure the {\em dc} voltage along the sample.  This is shown in Fig.~\ref{setup1} for a Py/Pt bilayer, with lateral dimensions of 2.92~mm~$\times$~20~$\mu$m and 15-nm thick individual layers.  The bilayer was prepared by optical lithography, sputter deposition, and lift-off on a GaAs substrate. Subsequently, we prepared Ag contacts for the voltage measurements, covered the whole structure with
100 nm of MgO (for {\em dc} insulation between bilayer and
waveguide), and defined a 30-$\mu$m wide and 200-nm thick Au
coplanar waveguide on top of the bilayer. Similar samples were
prepared with Pd, Au and Mo layers replacing Pt.

The high bandwidth of the coplanar waveguide setup enabled us to carry out measurements with microwave excitations in the frequency interval of 4--11~GHz. The power of the {\em rf} excitation was varied from 15 to 150~mW. For a given frequency, experiments were carried out as a function of external magnetic field $\vec{H}_{dc}$, with an in-plane orientation that could be rotated to arbitrary angles $\alpha$ with respect to the central axis of the coplanar wave guide.  While the FMR signal was determined from the impedance of the waveguide \cite{Mosendz-JAP08}, the {\em dc} voltage was measured simultaneously with a lock-in modulation technique as a function of $\vec{H}_{dc}$.

\subsection{FMR measurements}
\begin{figure}%
    \includegraphics[width=8.6cm]{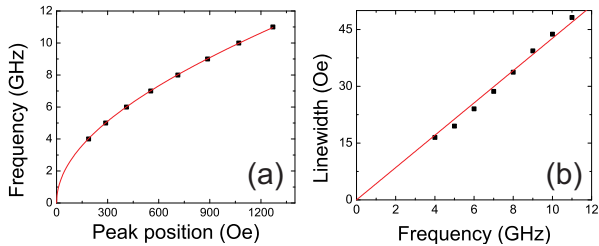}
    \caption{(Color online) Experimental data (symbols) for FMR peak positions and FMR linewidths of a Py/Pt bilayer are shown as a function of {\em rf} frequency in (a) and (b), respectively.  The solid line in (a) represents the fit to Eq.~(\ref{omega}) and results in $M_{s} = 852$~G.  The solid line in (b) represents a linear fit to the linewidth {\em vs.}\ frequency dependence.
    } %
    \label{Kittelandlw}%
\end{figure}

The FMR frequency {\em vs.}\ peak position for the Py/Pt sample is shown in Fig.~\ref{Kittelandlw}(a). Fitting the data to the Kittel formula:
\begin{equation}
  \left(\omega_f / \gamma_{g}\right)^{2}=H_{dc}(H_{dc}+4\pi M_{s}) \label{omega} ,
\end{equation}
results in the saturation
magnetization for Py of $M_{s} = 852$~G. Figure~\ref{Kittelandlw}(b) shows the FMR line width as a function of frequency. The linear behavior of the FMR line width indicates that damping in Py is governed by the intrinsic Gilbert phenomenology and any extrinsic effects are small.

\begin{figure}%
    \includegraphics[width=8.6cm]{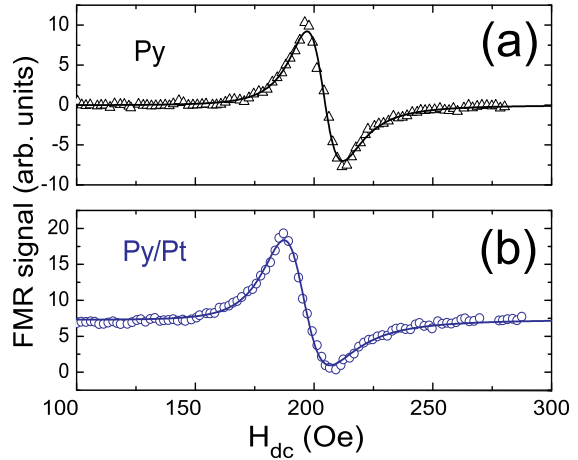}
    \caption{(Color online) Derivative FMR spectra at 4~GHz for (a) Py and (b) P/Pt.  Solid lines are Lorentzian line shape fits.
    } %
    \label{FMRspectra}%
\end{figure}

Figure~\ref{FMRspectra} shows FMR spectra for a Py/Pt bilayer and a Py single layer at 4-GHz excitation frequency. The FMR peak positions for the two samples are similar.  The main difference between the spectra is their FMR linewidth.  The FMR line widths (HWHM) extracted from fits
to Lorentzian absorption functions are $\Delta H_{Pt/Py} = 16.9$~Oe for Py/Pt and 
$\Delta H_{Py} = 12.9$~Oe for Py. The difference in $\Delta H$ can be attributed to the loss of pumped spin momentum in the Pt layer.  The thickness of the Pt layer is 15~nm, which is larger than $\lambda_{sd}^{Pt} =10 \pm 2$~nm \cite{Kurt-APL02}. Thus, all pumped spin momentum is dissipated in the Pt layer and we can extract the value of the spin mixing conductance 
$g^{\uparrow\downarrow}_{eff}$ from the increased linewidth. Using Eq.~(\ref{spinmix}) we calculate a spin mixing conductance $g^{\uparrow\downarrow}_{eff} = 2.1\times10^{19}$~m$^{-2}$ at the Py/Pt interface. This experimental value is somewhat smaller than  the previously reported 
$2.58 \times 10^{19}$~m$^{-2}$\cite{Mizukami-JMMM2001,Yaroslav-2002}. Cao \textit{et\ al.}\
\cite{Cao-JAP09} showed that for high power {\em rf} excitation, the
spin mixing conductance can be reduced due to the loss of coherent spin
precession in the ferromagnet.  This could be the case here, since the cone angle for the FMR at 4 GHz is relatively large, and a slightly larger mixing conductance for the smaller precession angles at 11 GHz would lead to more consistent frequency dependent values of the spin Hall angles as discussed below.

\subsection{{\em dc} Voltage due to ISHE and AMR effects}

\begin{figure}%
  \includegraphics[width=8.6cm]{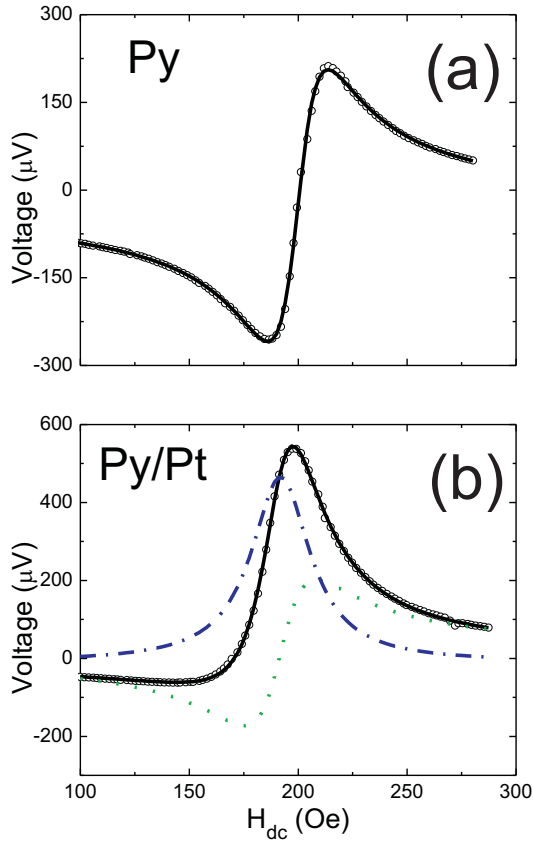}
  \caption{Voltage measured along the samples {\em vs.}\ field $H_{dc}$ for Py and Py/Pt at 4~GHz is shown with symbols in (a) and (b) respectively. Only the AMR contribution is present in the Py sample. Solid line in (a) shows a fit to Eq.~(\ref{AMR}). Both AMR and ISHE effects are observed in the Py/Pt. Dotted and dash-dotted lines show the AMR and ISHE contributions, respectively, which are extracted from fitting the data to Eqs.~(\ref{AMR}) and (\ref{ISHE}); the solid line in (b) shows the combined fit for the Py/Pt sample.} %
  \label{FMRSHE}%
\end{figure}

Figure~\ref{FMRSHE} shows the {\em dc} voltage measured along the samples with an external field applied at 45 degrees from the coplanar waveguide axis. For the Py/Pt sample we observe a resonant increase in the {\em dc} voltage along the sample at the FMR position.  However,
the lineshape is complex:  below the resonance field the voltage is negative, it changes sign just below the FMR resonance field, and has a positive tail in the high field region.  In contrast, the
single layer Py sample, which is not affected by spin pumping, shows a voltage signal that is purely antisymmetric with respect to the FMR position and thus mirrors the derivative FMR signal shown in Fig.~\ref{FMRspectra}(b).  The voltage generated by the ISHE depends only on the cone
angle $\theta$ of the magnetization precession [see
Eq.~(\ref{dcspincurrent})] and thus must be symmetric with respect
to the FMR resonance position. This means that the voltage measured in the Py/Pt sample has two contributions: (i) a symmetric signal due to the ISHE, and (ii) an antisymmetric signal of the same origin as that in the Py control sample.

\begin{figure}%
  \includegraphics[width=7cm,bb=0 0 320 490]{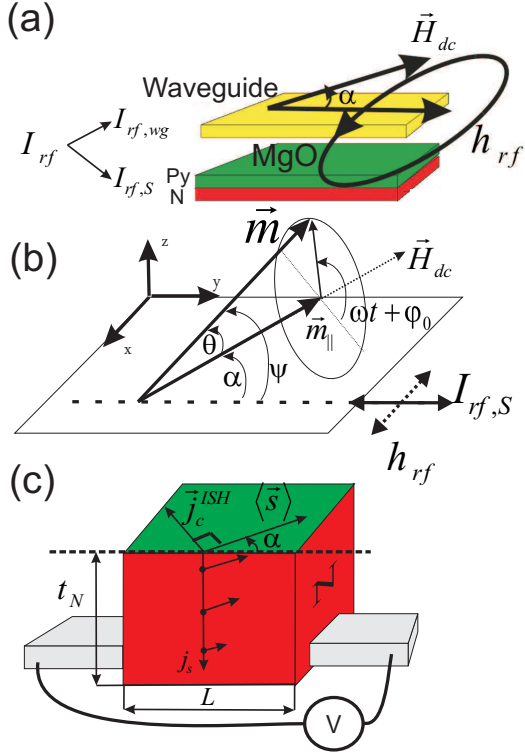}
  \caption{(Color online) Splitting of \textit{rf} current due to capacitive coupling is schematically shown in (a) together with the directions of the applied {\em dc} magnetic field $\vec{H}_{dc}$ and the {\em rf} driving field $\vec{h}_{rf}$ with respect to the bilayer and waveguide. (b) Schematic of 
 $\vec{m}$ precessing in Py. $\vec{m}$ precesses around its equilibrium direction given by 
 $\vec{H}_{dc}$ at the driving frequency $\omega_f$ and with a phase delay $\varphi_{0}$ with respect to $h_{rf}$.  $\alpha$ is the angle between $\vec{H}_{dc}$ and the waveguide axis (along $y$), $\theta$ is the cone angle described by $\vec{m}$ and $\psi$ is the angle between $\vec{m}$ and the waveguide axis. Due to the strong capacitive coupling part of  $I_{rf}$ flows through the Py given by $I_{rf,S}$. (c) Geometry of the \textit{dc} component of the pumped spin current with polarization direction $\langle\vec{s}\rangle$ along the equilibrium magnetization direction $\vec{m}_{||}$.  The charge current due to ISHE $\vec{j}_{c}^{ISH}$ is orthogonal to the spin current direction (normal to the interface) and $\langle\vec{s}\rangle$.  The voltage due to the ISHE is measured along $y$ (waveguide axis).  Solid arrows indicate the spin accumulation inside N, which decays with the spin diffusion length $\lambda_{sd}$.}
  \label{setup2}%
\end{figure}

The antisymmetric voltages observed in both Py and Py/Pt depend on the cone angle $\theta$ of the magnetization dynamics, since they vary rapidly around the FMR resonance position. This suggests that the antisymmetric signal originates from the 
AMR.\cite{Costache-APL06,Gui-APL2007,Gui-PRL2007}  Although the MgO provides
{\em dc} insulation between the sample and the waveguide, see Fig.~\ref{setup2}(a), the
strong capacitive coupling allows some leakage of the {\em rf} driving current 
$I_{rf} = I_{rf}^{m}\sin \omega_f t$ into the sample, $I_{rf,S}$, which flows along the waveguide direction.  Its magnitude can be estimated from the ratio between the waveguide resistance $R_{wg}$ and the sample resistance $R_S$: $I_{rf,S} = I_{rf}R_{wg}/R_S$, since the capacitive coupling impedance is negligible for $\omega_f \gtrsim 4$~GHz.  Furthermore, due to the strong capacitive coupling (50 pF) between sample and wave guide, both {\em rf} currents in the sample and waveguide are for all practical purposes in phase, {\em i.e.}, the relative phase shift is expected to be at most $10^{-3} \pi$.  Indeed, experiments with single layer Py 
samples\cite{Mosendz-PRL2010} and with a MgO layer inserted between the Py and Pt 
layers\cite{Mosendz-APL2010} are consistent with a pure AMR signal, as described below, without any appreciable phase shift.

The precessing magnetization in the Py [see Fig.~\ref{setup2}(b)] results in a time-dependent $R_S[\psi (t)] = R_{0} - \Delta R_{AMR}\sin^2\psi(t)$ due to the AMR given by $\Delta R_{AMR}$, where $R_{0}$ is the sample resistance with the magnetization along the waveguide axis, and $\psi$ is the angle between the instantaneous magnetization $\vec{m}$ and the waveguide axis [see Fig.~\ref{setup2}(b)]\cite{Costache-APL06}.  $\Delta R_{AMR}$ can be experimentally determined by static magnetoresistance measurements under rotation of an in-plane field sufficiently large to saturate the magnetization.  Since the AMR contribution to the resistance oscillates at the same frequency as the {\em rf} current, but phase-shifted, a homodyne
{\em dc} voltage develops and is given by:\cite{Mosendz-PRL2010}
\begin{equation}\label{AMR}
  V_{AMR}=I_{rf}^{m}\frac{R_{wg}}{R_{S}}
  \Delta R_{AMR}\frac{\sin(2\theta)}{2}\frac{\sin(2\alpha)}{2}\cos\varphi_{0} \; ,
\end{equation}
where $\varphi_{0}$ is the phase angle between magnetization precession and driving {\em rf} field, and the relation between $\theta$, $\alpha$ and $\psi$ is illustrated in
Fig.~\ref{setup2}(b).  Well below the FMR resonance the phase angle $\varphi_0$ is zero, it becomes $\pi /2$ at the peak, and is $\pi$ far above the resonance \cite{Bailey-JMMM06}.  Thus, 
$\cos\varphi_{0}$ changes sign upon going through the resonance, which gives rise to an
antisymmetric $V_{AMR}$, as is observed in both the Py and Py/Pt samples.
Following Guan {\em et al.}\ \cite{Bailey-JMMM06} we calculate
the cone angle $\theta$ and $\sin\varphi_{0}$ as a function of the
applied field $H_{dc}$, FMR resonance field $H_r$, FMR linewidth
$\Delta H$ and {\em rf} driving field $h_{rf}$:
\begin{equation}
\theta = \frac{h_{rf}\cos\alpha}{\Delta H\sqrt{1+
\left(\frac{(H_{dc}-H_{r})(H_{dc}+H_{r}+4\pi M_{s})}{\Delta H4\pi
M_{s}}\right)^{2}}} \label{teta}, \mathrm{and}
\end{equation}
\begin{equation}
\sin\varphi_{0} = \frac{1}{\sqrt{1+\left(\frac{(H_{dc}-H_{r})(H_{dc}+H_{r}+
4\pi M_{s})}{\Delta H4\pi M_{s}}\right)^{2}}} \label{phi0} \;.
\end{equation}
The anisotropic magnetoresistance was determined by {\em dc} magnetoresistance measurements with fields applied along the hard-axis as $\Delta R_{AMR} = 0.95$\%.  This allows us to fit the Py data [see Fig.~\ref{FMRSHE}(a)] with only one adjustable parameter $h_{rf}=4.5$~Oe using Eqs.~(\ref{AMR}--\ref{phi0}).

In order to understand the symmetric contribution to the Py/Pt voltage data we have to include
an additional voltage due to the ISHE.  In principle an inductive coupling (if any) could result in a symmetric voltage contribution to the signal.  However, this type of coupling is unlikely in our samples due to the fact that sample and transmission line are prepared as a stack with a thin insulator in the middle. Furthermore, our recent work~\cite{Mosendz-APL2010} showed that if spin pumping is suppressed by inserting a 3-nm MgO layer at the Py/Pt interface, then the symmetric part of the voltage vanishes. This unambiguously shows that the symmetric part of the measured voltage is related to spin accumulation in N, which appears due to the ISHE.  The absence of a symmetric contribution for Py alone also suggests that inductive effects are negligible.

In an open circuit, an electric field $\vec{E}$ is generated leading to a total current density
\begin{equation}\label{localcurrent}
\vec{j_{c}}(z)=j_{c}^{ISH}(z)(\hat{x}\cos\alpha+\hat{y}\sin\alpha)+\sigma\vec{E},
\end{equation}
where $\sigma$ is the charge conductivity and $\hat{x}$, $\hat{y}$ are defined in Fig.~\ref{setup2}. Since there is no current flowing in the open-circuit 
\begin{equation}\label{totalcurrent}
\int_{-t_{Py}}^{t_{N}}\vec{j_{c}}(z)dz=0 .
\end{equation}
When the wire is much longer than thick, the electric field in the wire is constant.  On the other hand, voltage generation occurs only in the Pt layer (more precisely in a skin depth of the spin-flip length in which the ISHE emf is generated), while the Py layer acts as a short, which decreases the voltage difference at the sample terminals.  Solving the system of Eqs.~(\ref{localcurrent}) and~(\ref{totalcurrent}), we obtain the component of the electric field along the
measurement direction $y$ as
\begin{equation}\label{efield}
E_{y}=-
\frac{Pg^{\uparrow\downarrow}_{eff}\sin\alpha\sin^{2}\theta\gamma e\omega_f\lambda_{sd}}
{2\pi(\sigma_{N}t_{N}+\sigma_{Py}t_{Py})}\tanh\left(\frac{t_{N}}{2\lambda_{sd}}\right),
\end{equation}
where $\sigma_N$ and $\sigma_{Py}$ are the charge conductivities in the N layer ({\em e.g.}, Pt) and Py, and $t_N$ and $t_{Py}$ are the thicknesses of the N and Py layers.  Using Eq.~(\ref{efield}) we calculate the voltage due to the ISHE generated along the sample with length $L$:
\begin{equation}\label{ISHE}
  V_{ISH}=-\frac{\gamma eLP\omega_f\lambda_{sd}g^{\uparrow\downarrow}_
  {eff}\sin\alpha\sin^{2}\theta}{2\pi(\sigma_{N}t_{N}+
  \sigma_{Py}t_{Py})}
\tanh\left(\frac{t_{N}}{2\lambda_{sd}}\right) .
\end{equation}
Note that this voltage is proportional to $L$ and, thus, sufficiently large voltage signals can be measured even for small $\gamma$ values by increasing the sample length.  Furthermore, note that for the case of the normal layer thickness $t_N$ being comparable to the spin diffusion length $\lambda_{sd}$ the measured voltage depends only very weakly on either value, since 
$t_N/\lambda_{sd}\tanh (t_N/2\lambda_{sd})$ is approximately constant.  

\begin{figure}%
    \includegraphics[width=8.6cm]{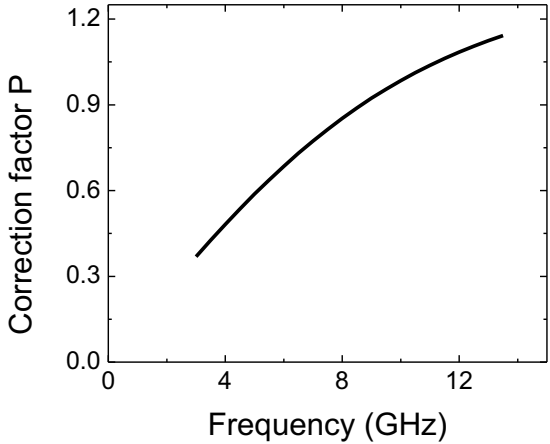}
    \caption{Elliptical precession trajectory results in a time-dependent cone angle of magnetization precession, that modifies the {\em dc} component of pumped spin current.  The ellipticity correction factor $P$ for the {\em dc} component of spin current is calculated as a function of microwave frequency according to Eq.~(\ref{corrfacelll}).
    } %
    \label{corrfac}%
\end{figure}

One of the input parameters in $V_{ISH}$ is the ellipticity correction factor $P$. At 4~GHz excitation, FMR occurs at $H_{dc} \approx 200$~Oe.  Therefore, the magnetization precession trajectory is highly elliptical and a correction to the {\em dc} voltage component due to the ellipticity is significant.   Figure~\ref{corrfac} shows $P$ as a function of microwave frequency as calculated using Eq.~(\ref{corrfacelll}). In the range from 4 to 13~GHz, $P$ changes almost by a factor of 3 and, therefore, has to be taken into account.  At frequencies above 10~GHz $P$ becomes larger than 1 and reaches a maximum value of 1.3 at $\approx 28$~GHz before it slowly decreases towards 1 for higher frequencies. This means that the most effective pumping of \textit{dc} component of spin current is achieved not for circular precession, but rather for some elliptical trajectory of magnetization precession. 

We used Eqs.~(\ref{ISHE}) and~(\ref{AMR}) to fit the voltage measured for the Py/Pt sample [see the solid line in Fig.~\ref{FMRSHE}(b)].  The dashed and dotted lines in Fig.~\ref{FMRSHE}(b) are the AMR and ISHE contributions, respectively.  By using a literature value for Pt of 
$\lambda_{sd} = 10 \pm 2$~nm \cite{Kurt-APL02}, the only remaining adjustable
parameters are the {\em rf} driving field $h_{rf}$ and the spin Hall angle 
$\gamma \approx 0.011 \pm 0.002$. Note that through the cone angle $\theta$, $h_{rf}$ affects both the AMR and ISHE contributions.  In fact, as seen from the fit to the control Py sample, $h_{rf}$ is already determined by the negative and positive tails of the AMR part.

\begin{figure}%
    \includegraphics[width=7cm]{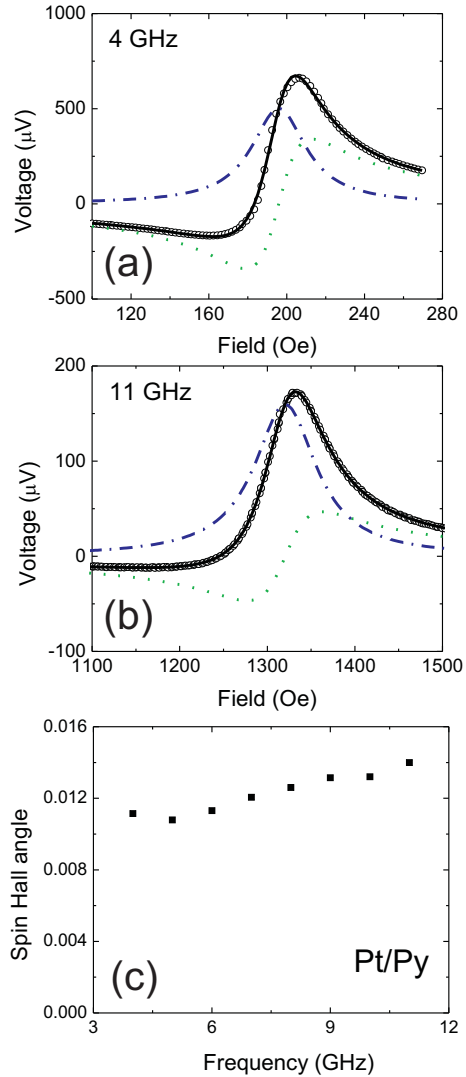}
    \caption{(Color online) (a) and (b) show voltages measured at 4 and 11 GHz. Voltages measured at 11 GHz are smaller due to a decreased precession cone angle. Note that the ratio between the ISHE and AMR contributions change due to a faster decrease of the AMR voltage at high frequencies.  (c) Spin Hall angle in Pt as a function of frequency. The slightly decreased values at lower frequencies may be due to non-linear effects and a concomitant decrease of spin pumping at large angles of magnetization precession.
    } %
    \label{shafreq}%
\end{figure}

We carried out additional measurements of the spin Hall angle as a function of the microwave frequency. Since the spin pumping is proportional to the time derivative of the magnetization [as manifested by the factor $\omega_f$ in Eq.~(\ref{ISHE})] and the ellipticity correction factor $P$, which increases with frequency (see Fig.~\ref{corrfac}), the voltages due to the ISHE are expected to increase at higher microwave frequencies. However, the cone angle of magnetization precession for a constant power of \textit{rf} excitation decreases due to higher resonance fields.  Since spin pumping is proportional to $\sin^{2}\theta$ an overall decrease in the voltage due to the ISHE is observed. However, the relative strength of the antisymmetric (AMR) and symmetric (ISHE) parts of the signal change since the AMR decreases faster than the ISHE contribution as a function of the frequency. This effect is illustrated in Figs.~\ref{shafreq}(a) and (b). 
Figure~\ref{shafreq}(c) shows the values of the spin Hall angle $\gamma$ extracted from the fits, which are essentially constant for all frequencies, except for a slight decrease of $\gamma$ at lower frequencies. Equation~(\ref{corrfacelll}) is strictly valid only for small precession cones and constant power of \textit{rf} excitation. From the fitting of the data we can extract the cone angles of the magnetization precession. At 4~GHz the fitted value $\theta \approx 10^{\circ}$ at the resonance, while at 11~GHz $\theta\approx 2.5^{\circ}$.  For 4-GHz excitation non-linear effects may start to play a role, possibly slightly changing the estimated value of $\gamma$.

\begin{figure}%
    \includegraphics[width=8.6cm]{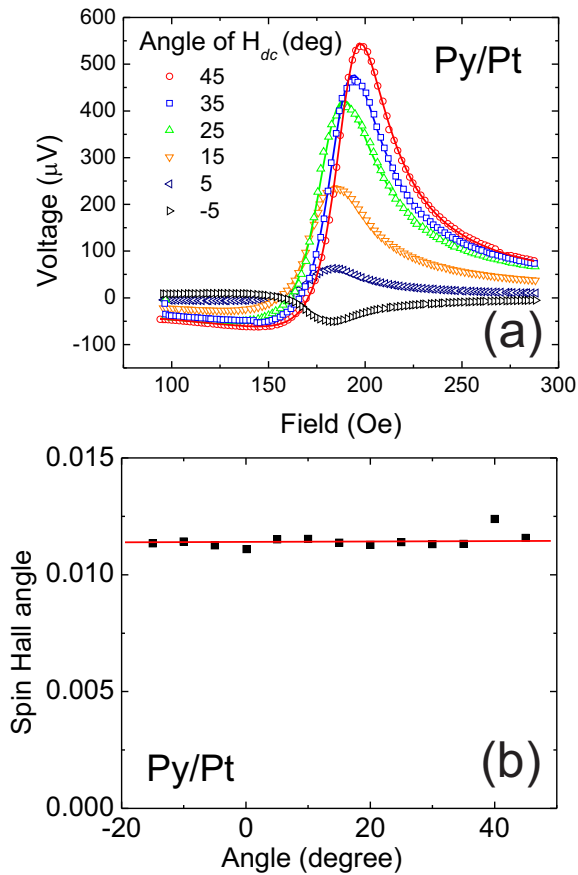}
    \caption{(Color online) (a) Voltage measured at 4~GHz as a function of angle $\alpha$ of the external magnetic field with respect to the coplanar waveguide axis. Experimental data and fits are shown with symbols and solid lines, respectively. (b) Spin Hall angle extracted from the fits.  The theoretical model correctly takes into account the angular dependence for the ISHE and AMR contributions.
    } %
    \label{shaang}%
\end{figure}

Our model was further tested by varying the angle $\alpha$ of the applied field with respect to the microwave transmission line.  Note that both Eqs.~(\ref{AMR}) and (\ref{ISHE}) have besides the explicit dependence on $\alpha$ an additional dependence through the implicit $\alpha$-dependence of $\theta$ given by Eq.~(\ref{teta}).  For small cone angles $\theta$ this results in both $V_{AMR}$ and $V_{ISH}$ being proportional to $\sin\alpha\cos^2\alpha$.  The dependence on the {\em dc} magnetic field direction is shown in Fig.~\ref{shaang}.  The measured voltage profile is consistent with the theoretical model, and results in a consistently constant fitted value of the spin Hall angle in Pt.  Due to the specific geometry of the sample we were not able to measure at angles close to $\alpha = 90^{\circ}$, at which the magnetization dynamics cannot be excited, because the component of $h_{rf}$ perpendicular to the magnetization vanishes and FMR cannot be excited.  But in the range of angles from $-5^{\circ}$ to $45^{\circ}$, excellent agreement between experiment and theory was achieved.

\begin{figure}%
    \includegraphics[width=8.6cm]{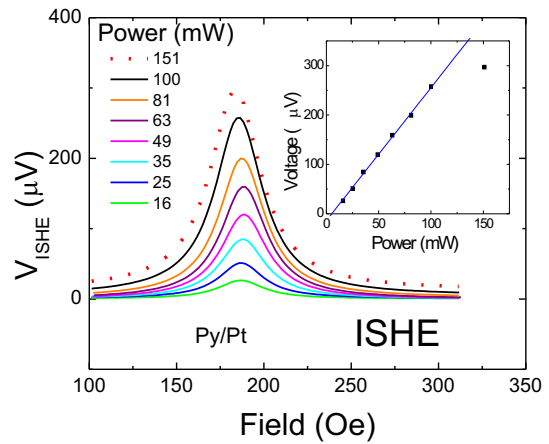}
    \caption{(Color online) Power dependence of the symmetric ISHE voltage contribution measured at 4~GHz. The inset shows that the maximum of the ISHE signal is linear {\em vs.}\ {\em rf} excitation power. The highest power (150~mW) deviates from the linear behavior due to excitation of non-uniform modes.  This deviation is also observed in the corresponding FMR spectra.
    } %
    \label{power}%
\end{figure}

The other adjustable parameter in our measurements is the power of the microwave excitation. The {\em rf} microwave field amplitude increases as a square root of the power. According to 
Eq.~(\ref{teta}) the cone angle $\theta$ of the magnetization precession increases linearly with driving field.  The voltage due to the ISHE is quadratic in $\theta$ and, thus, is expected to be proportional to the power.  After fitting the data, we extracted the symmetric part due to the ISHE, which is shown in Fig.~\ref{power}.  The maximum values of the measured voltage depend linearly on power, as expected by theory, except for the highest power of about 150~mW, at which the system is driven into the non-linear regime.  We observe a deviation of the FMR peak position as a well as a deviation of the FMR spectra from a Lorentzian shape. It is known that at high \textit{rf} power other modes beside the uniform FMR mode are excited. In this case one expects a substantial line broadening and even saturation of the FMR absorption, as observed.

\begin{figure}%
  \includegraphics[width=7.6cm]{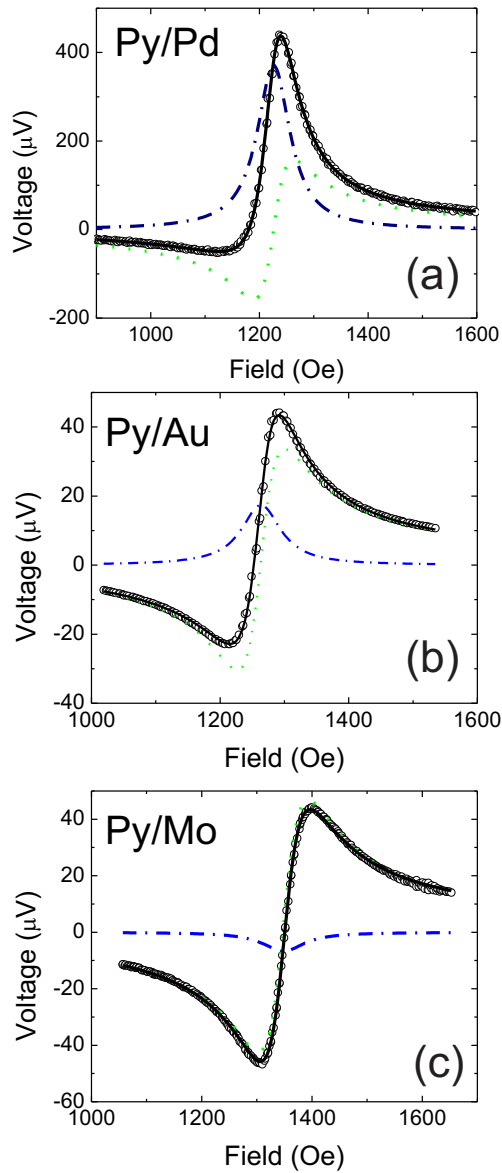}
  \caption{(Color online) Voltages measured at 11~GHz for (a) Py/Pd, (b) Py/Au, and (c) Py/Mo.  Shown are data (symbols), combined fits (black lines) and individual AMR and ISHE contributions, with dotted green and dash-dotted blue lines, respectively.   Note the opposite sign of the ISHE contribution for Mo compared to those for Au and Pd.} %
    \label{MoAu}%
\end{figure}

\begin{table}
\caption{Spin Hall angle $\gamma$ determined using $\lambda_{sd}$
and $\sigma_{N}$} from data measured at 11~GHz.
\begin{ruledtabular}
\begin{tabular}{cccc}\label{gammas}
Normal metal & $\lambda_{sd}$ (nm) & $\sigma_{N}$ $1/(\Omega m)$ & $\gamma$ \\
\hline
Pt   & 10$\pm$2 & (2.4$\pm$0.2)$\times10^{6}$ & 0.013$\pm$0.002 \\
Pd & 15$\pm$4 & (4.0$\pm$0.2)$\times10^{6}$ & 0.0064$\pm$0.001\\
Au  &  35$\pm$3 & (2.52$\pm$0.13)$\times10^{7}$ &  0.0035$\pm$0.0003 \\
Mo  & 35$\pm$3 & (4.66$\pm$0.23)$\times10^{6}$ & -0.0005$\pm$0.0001 \\
\end{tabular}
\end{ruledtabular}
\end{table}

\subsection{Spin Hall angle in Pd, Au and Mo}
Since the sample structure in our experiments is just a bilayer of Py with the non-magnetic material of interest, this technique can be readily applied to determine $\gamma$ in any
conducting material. In Fig.~\ref{MoAu} we show voltages measured for Py/Pd, Py/Au, and Py/Mo measured at 11~GHz excitation. The spin Hall contributions in Au and Mo are smaller than in Pt, and note that for Mo the spin Hall contribution has the opposite sign. Fitting of the data enabled us to extract the values of $\gamma$ for Pd, Au and Mo (see Table~\ref{gammas}).  The effective mixing conductance at the intermetallic Py/N interface is governed by N and we adopt the value obtained by the experimentally measured increased damping in Py/N.  Note that the determination of $\gamma$ furthermore requires $\sigma_{N}$ and $\lambda_{sd}$ as input parameters. 
$\sigma_{N}$ was obtained using four-probe measurements for all samples. Reported values for 
$\lambda_{sd}$ vary considerably.  We choose literature values for Pt and Pd from 
Ref.~\onlinecite{Kurt-APL02} and Au from Ref.~\onlinecite{Mosendz-PRB09}, and for Mo we assumed that $\lambda_{sd}$ is comparable to that for Au.  Even though this latter assumption may not necessarily hold, the sign change is consistent with earlier 
measurements~\cite{Morota-JAP09}.  We furthermore note that the $\gamma$ values in 
Table~\ref{gammas} differ from the previously reported ones in 
Ref.~\onlinecite{Mosendz-PRL2010}, where we assumed circular precession and therefore underestimated $\gamma$ by a factor of roughly 2.  In addition, Table~\ref{gammas} is based on 11~GHz data, which due to the smaller cone angles is less susceptible to deviations stemming from non-linear effects, and therefore should be more reliable. 

Our values for $\gamma$ are in good agreement with values reported by Otani {\em et al.}\ \cite{Vila-PRL07,Morota-JAP09} from measurements in lateral spin valves, but conflict with values reported by other groups\cite{Seki-NM2008,Ando-PRL2008}.  We note that in lateral spin
valves it is important to also understand the charge current distribution in order to rule out or correct for additional non-local voltage contributions \cite{Mihajlovic-PRL2009}.  A distinct advantage of our approach is that the measured voltage signal scales with the sample dimension, and no additional charge current is directly applied to the sample that could result in unwanted spurious voltage signals.

We also gain insights into spin orbit coupling in non-magnetic metals, which ultimately give rise to spin Hall effects.  Even for non-magnetic materials, that are next to each other in the periodic table (Pt and Au) the spin Hall angle differs almost by a factor of 4. On the other hand, Mo has a significantly smaller spin Hall angle with opposite sign.  This sign change can be rationalized by a simple $s$-$d$ hybridization model and is supported by first-principles 
calculations,\cite{Kotani-PRL2009} indicating that the spin Hall angle should be negative for less than half-filled $d$-bands, and positive for more than half-filled ones, consistent with our experimental results.  Pd in spite of being a lighter element than Au has a spin Hall angle which is almost 2 times larger.  First-principles calculations are again consistent with the experimental observation of $\gamma$ being larger for Pt and Pd compared to Au.\cite{Tanaka-PRB2008}  

\section{Conclusions}
We presented a spin pumping technique that enables measuring spin Hall angles in
various materials, which has clear advantages over standard {\em dc} electrical spin injection in Hall bar microstructures.  Our results for Pt, Pd, Au and Mo show that spin Hall angles are rather small, with the largest value found to be 0.013 for Pt.  Our approach provides a uniform spin current across a macroscopic sample.  The voltage signal from the inverse spin Hall effect can readily be increased via the use of longer samples, since $V_{ISH} \propto L$.  Furthermore, by using an integrated coplanar waveguide architecture we can control parameters, such as the {\em rf} driving field distribution, microwave frequency and power of \textit{rf} excitation.  This enabled a quantitative analysis of the data and a test of the theoretical model under various experimental conditions.  Our model accounts for both the anisotropic magnetoresistance and the inverse spin Hall effect contributions and agrees with experiments for a wide range of controllable parameters. We demonstrated the existence of symmetric (ISHE) and antisymmetric (AMR) voltages and could model the frequency, magnetic field direction, and excitation power dependence well. The AMR voltage in our experiments originates from capacitive coupling between the waveguide and the sample and is consistent with the parameters characteristic for the ferromagnetic resonance.  Our method will enable additional studies of spin Hall effects in other materials, and, therefore, will be useful to further understand the spin-orbit coupling mechanism in metals.  This is necessary in order to develop and optimize the spin Hall effect as a method to generate and detect spin currents in various circumstances, such as in the spin Seebeck effect\cite{Uchida-Nature2008}.

\section{Acknowledgements}
We would like to thank R.\ Winkler and G.\ Mihajlovi\'{c} for valuable discussions.  Furthermore, we would like to thank C.-M. Hu for pointing out that the angular dependence of the AMR and ISHE contributions are identical. This work was supported by the U.S. Department of Energy - Basic Energy Sciences under Contract No.\ DE-AC02-06CH11357 and EU-IST through project MACALO (no.~257159).

\bibliography{refer}

\end{document}